\documentclass[
    aps,
    prx,
    superscriptaddress,
    twocolumn,
    a4paper,
    floatfix,
    longbibliography
]{revtex4-2}

\usepackage{blindtext}
\usepackage{calc}
\usepackage{tikz}

\usepackage[american]{babel}
\usepackage[utf8]{inputenc}

\usepackage{grffile} 

\usepackage{newtxtext,newtxmath}
\usepackage{microtype}

\usepackage{amsmath}
\usepackage{amssymb}
\usepackage{bbm}
\usepackage{mathtools}
\usepackage{dsfont}
\usepackage{braket}
\usepackage{cancel}
\usepackage{slashed}
\usepackage{float}
\usepackage[scr=boondoxo]{mathalpha}

\usepackage{graphicx}

\usepackage{siunitx} 

\usepackage{ragged2e}
\usepackage{array}
\usepackage{tabularx}
\usepackage{booktabs}
\usepackage[export]{adjustbox}

\definecolor{cset-aps-blueberry}{RGB}{28,128,158}
\definecolor{cset-aps-blue}{RGB}{46,44,184}
\definecolor{cset-aps-turquoise}{RGB}{0,67,88}
\definecolor{cset-aps-limegreen}{RGB}{190,219,67}
\definecolor{cset-aps-green}{RGB}{31,138,112}
\definecolor{cset-aps-yellow}{RGB}{255,225,25}
\definecolor{cset-aps-orange}{RGB}{253,116,0}
\definecolor{cset-aps-red}{RGB}{219,0,43}

\makeatletter
\DeclareRobustCommand{\Arrow}[1][]{%
\check@mathfonts
\if\relax\detokenize{#1}\relax
\settowidth{\dimen@}{$\m@th\rightarrow$}%
\else
\setlength{\dimen@}{#1}%
\fi
\sbox\z@{\usefont{U}{lasy}{m}{n}\symbol{41}}%
\begin{picture}(\dimen@,\ht\z@)
\roundcap
\put(\dimexpr\dimen@-.7\wd\z@,0){\usebox\z@}
\put(0,\fontdimen22\textfont2){\line(1,0){\dimen@}}
\end{picture}%
}
\makeatother

\usepackage{hyperref}
\hypersetup{%
    colorlinks=true,
    linkcolor={cset-aps-red},
    linkbordercolor={cset-aps-red},
    filecolor={cset-aps-orange},
    filebordercolor={cset-aps-orange},
    citecolor={cset-aps-blue},
    citebordercolor={cset-aps-blue},
    urlcolor={cset-aps-green},
    urlbordercolor={cset-aps-green},
    menucolor={cset-aps-limegreen},
    menubordercolor={cset-aps-limegreen},
    breaklinks=true,
    pdfborderstyle={/S/U/W 2},
    pdfpagemode=UseOutlines,
    pdfstartpage={1},
}

\usepackage[noabbrev,capitalize]{cleveref}
\crefname{equation}{Eq.}{Eqs.}
\crefname{figure}{Fig.}{Figs.}
\crefname{table}{Tab.}{Tabs.}
\crefname{section}{Sec.}{Secs.}
\crefname{subsection}{Subsec.}{Subsecs.}
\crefname{subsubsection}{Subsubsec.}{Subsubsecs.}

\usepackage{lipsum}

\usepackage{placeins}
\usepackage{epsfig}
\usepackage[inline]{enumitem}
\setlist*[enumerate]{label=(\arabic*)}

\begin{document}

\title[Optimal joint cutting two-qubit rotation gates]{Optimal joint cutting of two-qubit rotation gates}
\collaboration{Published as
    \href{https://link.aps.org/doi/10.1103/PhysRevA.109.052440}
    {Phys.~Rev.~A,  \textbf{109}, 052440 (2024)}}
\author{Christian Ufrecht}
\email{christian.ufrecht@iis.fraunhofer.de}
\address{Fraunhofer IIS, Fraunhofer Institute for Integrated Circuits IIS, Division Positioning and Networks, Nuremberg, Germany}
\author{Laura S. Herzog }
\address{Fraunhofer IIS, Fraunhofer Institute for Integrated Circuits IIS, Division Positioning and Networks, Nuremberg, Germany}
\author{Daniel D. Scherer }
\address{Fraunhofer IIS, Fraunhofer Institute for Integrated Circuits IIS, Division Positioning and Networks, Nuremberg, Germany}
\author{Maniraman Periyasamy}
\address{Fraunhofer IIS, Fraunhofer Institute for Integrated Circuits IIS, Division Positioning and Networks, Nuremberg, Germany}
\author{Sebastian Rietsch}
\address{Fraunhofer IIS, Fraunhofer Institute for Integrated Circuits IIS, Division Positioning and Networks, Nuremberg, Germany}
\author{Axel Plinge }
\address{Fraunhofer IIS, Fraunhofer Institute for Integrated Circuits IIS, Division Positioning and Networks, Nuremberg, Germany}
\author{Christopher Mutschler }
\address{Fraunhofer IIS, Fraunhofer Institute for Integrated Circuits IIS, Division Positioning and Networks, Nuremberg, Germany}

\begin{abstract}
    Circuit cutting, the partitioning of quantum circuits into smaller independent fragments, has become a promising avenue for scaling up current quantum-computing experiments. Here, we introduce a scheme for joint cutting of two-qubit rotation gates based on a virtual gate-teleportation protocol. By that, we significantly lower the previous upper bounds on the sampling overhead and prove optimality of the scheme. Furthermore, we show that no classical communication between the circuit partitions is required. For parallel two-qubit rotation gates we derive an optimal ancilla-free decomposition, which include CNOT gates as a special case. 
\end{abstract}

\maketitle

\section{Introduction}
Current quantum computing hardware faces serious limitations, such as low qubit numbers and high susceptibility to noise. As a result, quantum hardware will likely be unable to execute algorithms with provable speedup like Shor's \cite{Shor1994} or Grover’s \cite{Grover1996} algorithm in the near future. 
On the other hand, experiments with heuristic quantum algorithms have been conducted on a small number of qubits, primarily applied to toy problems from the fields of, e.g.~optimization \cite{Harrigan2021,Farhi2014}, machine learning \cite{Havlivcek2019,Biamonte2017}, and chemical simulations \cite{Lanyon2010,Kandala2017}. However, scaling up both problem sizes and the number of qubits is crucial for assessing the potential usefulness of quantum computers. Even though recent progress \cite{Kim2023} suggests that quantum hardware based on superconducting qubits might be on the verge of entering a so-called \textit{utility regime} where the first practically interesting problems might be approached, to achieve relevant hardware size, modular quantum computing has been proposed \cite{Devoret2013,Monroe2013}. In this paradigm, multiple quantum computing platforms are interconnected through quantum links possibly enhanced by virtual entanglement distillation \cite{Yuan2023,bechtold2023}. Until reaching a higher technology readiness level, circuit cutting could serve as a useful approach, replacing quantum links with classical links and a post-processing step.

Given a unitary quantum channel $\mathcal{V}$, circuit cutting describes the method of decomposing the channel as  
\begin{equation}
\label{generalDecomposition}
    \mathcal{V}=\sum_i a_i \mathcal{F}_i\,,
\end{equation} 
with the real coefficients $a_i$, to reduce the circuit size or the impact of noise. 

Circuit cutting can be categorized into two methods. \textit{Wire cutting} \cite{Peng2020, Brenner2023, Harada2023,Lowe2022,Uchehara2022,Pednault2023} effectively corresponds to the cutting of horizontal empty qubit wires which are modeled mathematically by the identity channel. Consequently, wire cutting is a decomposition of this channel into measure-and-prepare channels. The second method, on which the focus will be in this work, is known as \textit{gate cutting} \cite{Hofmann2009,Mitarai2021,Mitarai2021b, Piveteau2022_circuitcut,Ufrecht2023}. Here,  $\mathcal{F}_i$ are elements of LOCC($A$,$B$), that is local operations on two partitions $A$ and $B$ of the quantum circuit and classical communication between them. Since $\mathcal{F}_i$ are local operations, no entanglement is created by the channel, but is effectively simulated as described below. If all wires or gates that connect the partitions $A$ and $B$ are cut, the sub-circuits become independent and can be run on two quantum computers only connected by classical communication links. Most circuit-cutting research so far has focused on applications to algorithms where the output is an expectation value of an observable. To evaluate \cref{generalDecomposition} in this context, we first define $\kappa=\sum_i |a_i|$  and the probability distribution  $p_i=|a_i|/\kappa$. Next, \cref{generalDecomposition} is rewritten as 
\begin{equation}
\label{QPD}
    \mathcal{V}=\kappa \sum_i  p_i\mathrm{sign}(a_i) \mathcal{F}_i
\end{equation}
\begin{figure}[t]
    \centering\includegraphics[width=246pt]{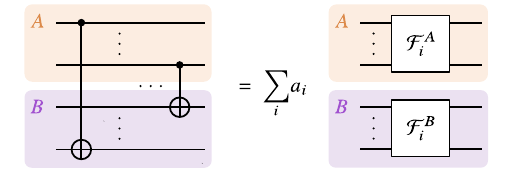}
    \caption{Cutting of parallel CNOT gates. Consider two partitions $A$ and $B$ of a quantum circuit connected by $n$ CNOT gates. In the case of gates that can be executed at the same time slice of the circuit (parallel gates), we present a joint optimal ancilla-free decomposition with coefficients $a_i$ and unitary or measurement operations $\mathcal{F}^A_i$ and $\mathcal{F}^B_i$. The sampling overhead is $\mathcal{O}(\gamma^2)$ where $\gamma=\mathcal{O}(2^n)$.   This is a strong improvement compared to cutting $n$ CNOT gates independently for which $\gamma=\mathcal{O}(3^n)$. Note that the equality in the figure has to be understood on the superoperator rather than on the gate
level.}\vspace*{-9pt}
    \label{fig:parallel_gates}
\end{figure}
and evaluated via Monte-Carlo sampling. At each experimental shot we select the $i$th channel $\mathcal{F}_i$ with probability $p_i$ and evaluate the circuit with $\mathcal{V}$ replaced by $\mathcal{F}_i$. The measurement outcome is weighted by $\kappa \mathrm{sign}(a_i)$ and the mean over many runs produces an unbiased estimate for the expectation value of the observable. This sampling procedure is referred to as quasi-probability sampling \cite{Pashayan_2015} because of the appearance of the sign of $a_i$ in \cref{QPD} which is referred to as a quasi-probability decomposition (QPD).

Quasi-probability sampling incurs a sampling overhead, the factor of more samples required to estimate the expectation value of an observable with the same accuracy as with respect to the original uncut circuit. This sampling overhead has been shown to be $\kappa^2$ \cite{Pashayan_2015, Piveteau_2022_quasiprob} and originates from the factor of $\kappa$ in front of the sum in \cref{QPD}.

Note the similarity of the approach to probabilistic error cancellation where the (possibly unphysical) inverse of a noisy quantum channel is decomposed into physical operations \cite{Temme_2017}.

The application of circuit cutting to sampling tasks has been studied initially in Ref. \cite{Lowe2022} and recently more extensively in Ref. \cite{Herzog2024} with the result that circuit cutting can also be meaningful in this situation.

When gates or wires are cut individually, the overall sampling overhead increases exponentially in the number of gates and wires cut. This strong increase of sampling overhead renders circuit cutting prohibitively expensive when the cutting location in a circuit and the cutting scheme is not carefully selected.
The task is, therefore, to  find decompositions of gates or collections of gates with minimal $\kappa$, which we will refer to as $\gamma$.

Ref. \cite{Piveteau2022_circuitcut} employed CNOT-gate teleportation \cite{Collins2001,Gottesman1999, Eisert2000} consuming one Bell state initially connecting the two partitions per CNOT gate. The optimal QPD with $\gamma=3$ of the Bell-state density matrix then translates into the optimal decomposition of the gate. 
The crucial insight of Ref.~\cite{Piveteau2022_circuitcut} was that the joint QPD of $n$ Bell states required for the teleportation of $n$ CNOT gates can be constructed more efficiently compared to the individual decomposition of the gates. As a result, the $\gamma$ parameter is reduced from $\mathcal{O}(3^{n})$ to $\mathcal{O}(2^{n})$, however at the cost of one ancilla qubit per partition and gate. In this cutting scheme, $\mathcal{F}_i\in \mathrm{LOCC}(A,B)$ and two-way classical information is shared between the partitions.  The protocol was extended to general Clifford gates and later also to wire cutting \cite{Brenner2023}. Recently, it was shown that optimal cutting of parallel wires is possible without ancilla qubits \cite{Harada2023}.

In this work, we extend previous proposals to the joint cutting of non-Clifford two-qubit rotation gates. We significantly simplify existing methods and, by that, enable implementation on current noisy intermediate-scale quantum (NISQ) hardware. Two-qubit rotation gates play a crucial role for example in the quantum approximate optimization algorithm (QAOA) \cite{Farhi2014} or for the simulation of spin systems.
In \cref{Virtual gate teleportation}, we introduce a virtual teleportation scheme that effectively consumes less than one entanglement bit (ebit). After explaining the protocol for a single gate instance, we generalize the scheme to the joint cutting of $n$ two-qubit rotation gates and demonstrate the optimality of the derived circuit-cutting method. With optimality, we refer to the circuit cutting scheme based on quasi-probability sampling with the minimal possible sampling overhead.
A two-qubit rotation gate with rotation angle $\theta$ is defined as
\begin{equation}
    R_{zz}(\theta)=\mathrm{cos}(\theta/2)\mathbb{I}\otimes \mathbb{I} -i\mathrm{sin}(\theta/2) Z\otimes Z
\end{equation}
with the single-qubit identity $\mathbb{I}$ and the Pauli $Z$ matrix.
The $\gamma$ parameter for this gate is \cite{Mitarai2021, Piveteau2022_circuitcut}
\begin{align}
\label{gammaRZZ}
\gamma&=1+2|\sin(\theta)|\,.\\
\intertext{
As we will show, when $n$ instances of this gate are jointly cut, we will find the reduced effective $\gamma$ parameter per gate}
\gamma&=1+|\sin(\theta)|
\end{align} 
asymptotically as $n$ approaches infinity, which matches the lower bound. 
Our scheme offers several improvements over previous work. First, it eliminates the need for classical communication between the partitions, and consequently, no real-time feedback is required. This contrasts the gate-teleportation protocol used in Ref.~\cite{Piveteau2022_circuitcut} where local correction operators conditioned on intermediate measurement results are necessary. 
Moreover, as explained in more detail in \cref{Virtual gate teleportation}.B our protocol can be executed sequentially on the same quantum computer, unlike schemes where two-way classical communication exchange necessitates the use of two separate quantum computers in parallel. Second, it has recently been emphasized that the number of operations $\mathcal{F}_i$ can be the limiting factor for the execution of circuit-cutting protocols \cite{Harada2023}.
We derive a quasi-probability decomposition of the entanglement resource states with an exponentially reduced number of elements. 
Third, in \cref{Parallel gates can be cut without ancilla qubits}, we show that parallel gates can be cut without the need for ancilla qubits. Parallel gates refer to gates that can be executed within the same time slice of a circuit as shown in \cref{fig:parallel_gates}.
Note that controlled rotation gates and CNOT gates (for $\theta=\pi/2$) are equivalent to 
two-qubit rotation gates up to local operations. Thus, all results of our work also apply to these types of gates.

\section{Virtual gate teleportation}
\label{Virtual gate teleportation}
\begin{figure}[bt]
    \centering
   \includegraphics[width=246pt]{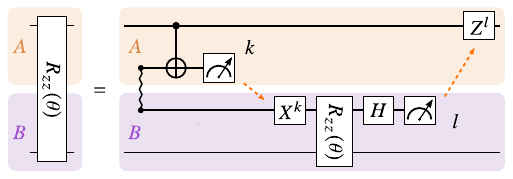}
   \caption{Teleportation protocol for a two-qubit rotation gate $R_{zz}(\theta)$ \cite{Eisert2000}. The scheme consumes an initial Bell state (wiggly line) as entanglement resource between the partitions $A$ and $B$ and requires two-way classical communication visualized by the dashed arrows. The measurements are performed in the computational basis with the outcomes $l,k\in\{0,1\}$. Here, $H$ is the Hadamard gate and $Z$ and $X$ are the Pauli operators}. 
   \label{fig:teleportation}
\end{figure}

In this section, we aim to find an optimal joint decomposition of two-qubit rotation gates based on gate teleportation.
A teleportation scheme for a two-qubit rotation gate $R_{zz}(\theta)$ \cite{Eisert2000} is shown in \cref{fig:teleportation}. It consumes one initial Bell state (wiggly line in the figure) prepared on two ancilla qubits. These qubits are subsequently measured in the computational basis, and the outcomes, $k,l\in\{0,1\}$, are communicated between the partitions $A$ and $B$ via classical communication to select the local Pauli correction operators $X^k$ and $Z^l$. For $\theta=\pi/2$ the two-qubit rotation gate is equivalent to a CNOT gate up to local operations.
In this case, simulating the initial Bell state by a quasi-probability distribution with $\gamma=3$ translates into an optimal cutting scheme for a CNOT gate (also $\gamma=3$) \cite{Piveteau2022_circuitcut}. When employing the teleportation scheme for $n$ CNOT gates together with a joint quasi-probability decomposition of the $n$ Bell states, the $\gamma$ parameter behaves sub-multiplicatively reducing from $\mathcal{O}(3^n)$ to $\mathcal{O}(2^n)$.

When attempting to achieve a similar reduction for joint cutting of two-qubit rotation gates, we encounter the following difficulty: A decomposition of a two-qubit rotation gate based on the teleportation scheme would be sub-optimal since it always consumes one Bell state resulting in $\gamma=3$ while the $\gamma$ parameter of a two-qubit rotation gate in \cref{gammaRZZ} is smaller for most values of $\theta$.
On the other hand, a deterministic gate teleportation protocol consuming less than one ebit does not exist  \cite{Stahlke2011,Soeda2011} but only probabilistic implementations \cite{Cirac2001, Dur2001,Groisman2005}, which would lead to suboptimal sampling overhead. As a result, a naive generalization of Ref.~\cite{Piveteau2022_circuitcut} is impossible.

\subsection{Single gate instance}
\begin{figure}[bt]
    \centering\includegraphics[width=246pt]{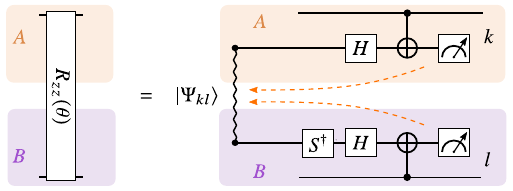}
    \caption{Virtual gate teleportation. As explained in the main text, for the state shown in \cref{state}, the correct gate is only teleported for the measurement results $k=l$, resulting in suboptimal sampling overhead when discarding other outcomes. Since we are only interested in a teleportation scheme together with a quasi-probability simulation of the initial state, we show that in this case, the result $k \neq l$ can be corrected in the final classical post-processing step.
This effectively corresponds to the preparation of the initial state $|\Psi_{kl}\rangle$ defined in \cref{reduced Bell state}, which depends on the outcome of the measurement performed later in time, therefore the term \textit{virtual teleportation}. In the figure, $S$ is the phase gate.}  
    \label{fig:virt_teleportation}
\end{figure}

In the following, we introduce the alternative scheme depicted in \cref{fig:virt_teleportation} for which we prepare the initial state $|\Psi_{kl}\rangle$, to be specified below, on the ancilla qubits.
As a first step, we choose the initial state
\begin{equation}
\label{state}
    |\Psi\rangle =\mathrm{cos}(\theta/2)|00\rangle+\mathrm{sin}(\theta/2)|11\rangle\,.
\end{equation}
 As we detail in \cref{Optimal QPD for a pure-sate denisty martrix}, an optimal QPD of a pure-state density matrix can be readily calculated based on the robustness of entanglement measure \cite{Vidal_1999, Piveteau2022_circuitcut}. The $\gamma$ parameter is calculated as
\begin{equation}
\label{calculation gamma}
    \gamma= 2\Big(\sum_j c_j\Big)^2-1
\end{equation} 
where $c_j$ are the Schmidt-coefficients of the state. Consequently, for the state shown in \cref{state}, we obtain $\gamma=2(|\cos(\theta/2)|+|\sin(\theta/2)|)^2-1$ matching \cref{gammaRZZ}.
However, it is straightforward to convince ourselves that the correct gate is teleported only for $k=l$, and its Hermitian conjugate for $k\neq l$ \cite{Dur2001}. On the one hand, local operators on the partitions $A$ and $B$ conditioned on the measurement results to achieve a deterministic protocol do not exist \cite{Stahlke2011,Soeda2011}. On the other hand, a \textit{repeat-until-success} version that is disregarding all measurement outcomes with $k\neq l$ would lead to the suboptimal sampling overhead $4\gamma^2$.
Note, however, that we are not interested in deriving an actual deterministic gate teleportation protocol but in a QPD to simulate the two-qubit rotation gate with optimal sampling overhead. 

In \cref{Optimal QPD for a pure-sate denisty martrix}, we derive an alternative optimal QPD for a pure-state density matrix compared to the one presented in Ref.~\cite{Vidal_1999}. Applied to \cref{state}, we find
\begin{align}
 \label{decomposition_single_gate3}
|\Psi\rangle\langle\Psi|&=\mathrm{cos}^2(\theta/2)|00\rangle \langle00|+\mathrm{sin}^2(\theta/2)|11\rangle \langle11|\\
  \label{decomposition_single_gate4}
 &+\mathrm{sin}(\theta) (\sigma^+-\sigma^-)\,.
\end{align}
The explicit form of the separable states $\sigma^+$ and $\sigma^-$ is specified in \cref{Optimal QPD for a pure-sate denisty martrix}.
To see that the decomposition is optimal, we sum over the absolute of the prefactors, finding again \cref{gammaRZZ}. The states $\sigma^+$ and $\sigma^-$ are independent of $\theta$, which is the crucial property to perform our protocol as described in the following:
 As mentioned above, for $k\neq l$, the Hermitian conjugate of the two-qubit rotation gate is teleported, which corresponds to $\theta\rightarrow-\,\theta$ in \cref{decomposition_single_gate3} and \cref{decomposition_single_gate4}. Because
 $\sin(-\,\theta)=-\sin(\theta)$, we could equally choose a minus sign in front of \cref{decomposition_single_gate4}, keep results with $k\neq l$, and disregard results with $k=l$.
\begin{figure*}
    \centering
   \includegraphics[width=510pt]{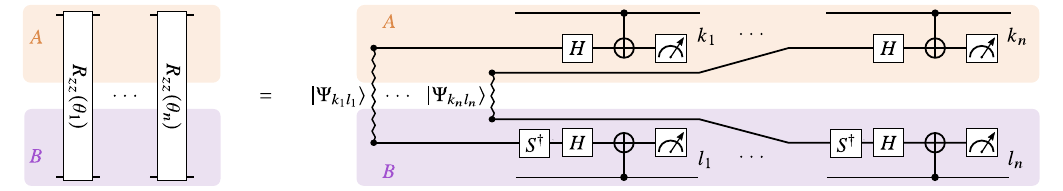}
   \caption{Generalization of the virtual teleportation protocol from the single gate instance shown in \cref{fig:virt_teleportation} to $n$ two-qubit rotation gates. Each gate requires one ancilla qubit per partition on which we effectively prepare the states $|\Psi_{k_sl_s}\rangle$ for $s=1,...,n$ by a quasi-probability decomposition. The result of the measurements on the ancilla qubits determines the sign of the quasi-probabilities in the classical post-processing step.} 
   \label{fig:teleportation_many}
\end{figure*}
To make use of both situations, we define the QPD
  \begin{align}
 \label{decomposition_single_gate1}
|\Psi_{kl}\rangle\langle\Psi_{kl}|&=\mathrm{cos}^2(\theta/2)|00\rangle \langle00|+\mathrm{sin}^2(\theta/2)|11\rangle \langle11|\\
  \label{decomposition_single_gate2}
 &+(-1)^{k+l}\mathrm{sin}(\theta) (\sigma^+-\sigma^-)
\end{align}
in which the signs of the quasi-probabilities depend on the measurement outcomes in the form of the factor $(-1)^{k+l}$. To execute the protocol, we proceed as detailed in the following: To estimate an expectation value experimentally, one typically defines a post-processing function $f(s)\in[-1,1]$ \cite{Peng2020} as a function of the bitstring $s$ observed at each experimental run and then takes the sample mean over many runs. As explained in the introduction,  we perform Monte-Carlo sampling as, for example, explained in Ref.~\cite{Ufrecht2023}. For each run, we prepare one of the states in \cref{decomposition_single_gate1} and \cref{decomposition_single_gate2} sampled according to the probability given by the absolute of the prefactor divided by $\gamma$. In case we initialize the ancilla qubits with one of the states constituting $\sigma^+$ or $\sigma^-$, we accumulate $\pm(-1)^{k+l}\gamma f(s)$ in the sample mean where the sign depends on the outcome of the measurements on the ancilla qubits.
We will refer to our protocol as \textit{virtual gate teleportation} since an actual entanglement resource state is not physically prepared but only simulated by a quasi-probability decomposition. In this respect, the protocol effectively makes use of the state
\begin{equation}
\label{reduced Bell state}
    |\Psi_{kl}\rangle =\mathrm{cos}(\theta/2)|00\rangle+(-1)^{k+l}\mathrm{sin}(\theta/2)|11\rangle
\end{equation}
which depends on the outcomes of the measurement performed later in time.

\subsection{Joint cutting of multiple gates}
\label{Joint cutting of multiple gates}
In this subsection, we generalize the virtual protocol to the joint cutting of $n$ two-qubit rotation gates at arbitrary positions in the circuit.
When cutting $n$ gate instances independently, the overall $\gamma$ parameter defined here as $\gamma^{(n)}_\mathrm{ind.}$ behaves multiplicatively, that is 
\begin{equation}
    \gamma^{(n)}_\mathrm{ind.}=\Pi_{s=1}^n (1+2|\mathrm{sin}(\theta_s)|)
\end{equation}
where we used \cref{gammaRZZ}.
In order to investigate if joint decomposition can be done more efficiently, the single-instance virtual teleportation protocol is generalized in the following to $n$ gates as shown in \cref{fig:teleportation_many}.
 In the figure we view the $n$ two-qubit states $|\Psi_{k_sl_s}\rangle$ for $s=1,...,n$ on the ancilla qubits as a $2n$-qubit state $|\Psi_{k_1l_1...k_nl_n}\rangle$.  Again, a specific choice of $k_s, l_s$ for $s=1,...,n$ defining the resource state only translates into the signs of the quasi probabilities of its QPD. In turn, we can choose $k_s=l_s=0$ for all $s$ of the resource state but correct unwanted measurement results in the post-processing step.
 
 Using \cref{reduced Bell state} and the qubit arrangement as shown in \cref{fig:teleportation_many}, we find
\begin{equation}
\label{decomposition_n_state}
|\Psi_{k_1l_1...k_nl_n}\rangle=\sum_{j\in\{0,1\}^n}c_j |j_1...j_n\rangle| j_n...j_1\rangle
\end{equation}
with $c_j=\Pi_{s=1}^n\mathscr{c}_{j_s}$
and
\begin{equation}
\label{coefficients joint}
\mathscr{c}_{j_s} = \left\{\begin{array}{lr}
        \mathrm{cos}(\theta_s/2) & \quad\text{for } j_s=0\\
        (-1)^{k_s+l_s}\,\mathrm{sin}(\theta_s/2) &\quad \text{for } j_s=1
        \end{array}\right.\,.
\end{equation}
In \cref{decomposition_n_state}, the first (second) ket describes the state on the ancilla qubits of partition $A$ ($B$).
If we redefine, e.g.,~the first ket on the right-hand side of \cref{decomposition_n_state} by absorbing the sign of $c_j$, this equation is the Schmidt decomposition of $|\Psi_{k_1l_1...k_nl_n}\rangle$ allowing us to calculate $\gamma_{\mathrm{joint}}^{(n)}$ for the QPD as
\begin{align}
    \label{derivation_gamma_1}
\gamma_{\mathrm{joint}}^{(n)}&=2\Big(\sum_{j\in\{0,1\}^n}|c_j|\Big)^2-1\\
\label{derivation_gamma_3}
    &=2\,\Pi_{s=1}^n(1+|\mathrm{sin}(\theta_s)|)-1\\
    \label{derivation_gamma_2}
    &< \gamma_{\mathrm{ind.}}^{(n)}\,.
\end{align}
In \cref{Optimal QPD for a pure-sate denisty martrix}, we explicitly state an optimal quasi-probability decomposition for any pure-state density matrix. Applied to \cref{decomposition_n_state}, the measurement-dependent signs in \cref{coefficients joint} only appear as the signs of the quasi probabilities, again allowing correction in the classical post-processing step. 
The $\gamma$ parameter in \cref{derivation_gamma_3} is, in general, significantly smaller than the $\gamma$  parameter for independent cuts. Therefore, we have shown that there exists a decomposition for multiple instances of two-qubit rotation gates with sub-multiplicative behavior of the $\gamma$ parameter. In \cref{Lower bounds}, we prove that $\gamma_{\mathrm{joint}}^{(n)}$ also is a lower bound for the optimal decomposition of $n$ two-qubit rotation gates. Consequently, joint optimal virtual gate teleportation also translates into optimality of the gate decomposition.

Note that no classical communication is required between the partitions \footnote{It can be shown that all operations in the virtual teleportation protocol are elements of the extended definition of local operations used in Ref.~\cite{Piveteau2022_circuitcut, Brenner2023}.}. Joint gate cutting without the need for classical communication significantly alleviates implementation on current hardware for two reasons. First, no real-time feedback for local operations conditioned on measurement results is needed, distinct from conventional gate teleportation protocols as, for example, shown in \cref{fig:teleportation}.
Second, if all gates connecting the two partitions are cut, the severed sub-circuits can be executed sequentially on the same hardware.
Indeed, given a decomposition as in \cref{generalDecomposition}, we first determine the number of shots $N$ required to achieve a certain accuracy of the expectation value estimate with high probability. Next, we determine how many times the circuit with channel $\mathcal{F}_i$ needs to be executed by sampling $N$ times from the probability distribution with $p_i=|a_i|/\kappa$ and counting the number of times $i$ is realized. 
If the decomposition does not require classical communication, $\mathcal{F}_i=\mathcal{F}_i^{A}\otimes \mathcal{F}_i^{B}$ and if $\mathcal{F}_i$ on one partition involves measurements, the $i$th channel on the other partition is independent of the measurement result. If all gates connecting the two partitions are cut, the circuits can be evaluated sequentially and independently from each other since $\mathcal{F}_i^{A}$ and $\mathcal{F}_i^{B}$ are completely independent. After all circuits have been executed, the bitstrings that were measured on both partitions for the $i$th channel are concatenated and passed to the post-processing function.
This procedure stands in stark contrast to protocols employing two-way classical communications where two quantum devices connected by classical links must operate in parallel. This results from the structure of $\mathcal{F}_i$ when allowing classical communication. Here, we allow operations that may depend on the outcome of a measurement on the other partition. Since measurement results are probabilistic, the number of times each channel has to be applied cannot be calculated prior to the experiment.

Lastly, we mention again that our protocol contains the joint cutting of $n$ CNOT gates as a special case with optimal $\gamma_\mathrm{joint}^{(n)}=2^{n+1}-1$ \cite{Piveteau2022_circuitcut}.

\subsection{Multi-qubit rotation gates}
\begin{figure}[H]
    \centering\includegraphics[width=246pt]{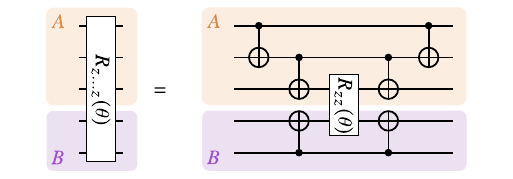}
    \caption{A multi-qubit rotation gate is equivalent to a two-qubit rotation gate up to local operations (with respect to the two-partitions $A$ and $B$). Consequently, the $\gamma$ parameter of this gate equals the one of the two-qubit rotation gate.}
    \label{fig:multi qubit}
\end{figure}

Finally, we consider multi-qubit $Z$ rotation gates. A multi-qubit rotation gate can be written as a two-qubit rotation gate sandwiched by ladders of CNOT gates \cite{Cowtan2020} as shown in \cref{fig:multi qubit}. The task of cutting a multi-qubit $Z$ rotation gate then reduces to the cutting of a two-qubit rotation gate. 
The resulting decomposition is optimal. This can be seen by noting that if the $\gamma$ parameter of the multi-qubit rotation gate was smaller than that of a two-qubit rotation gate, we could use the former to construct a decomposition of the latter with smaller $\gamma$, leading to a contradiction.

\section{Parallel gates can be cut without ancilla qubits}
\label{Parallel gates can be cut without ancilla qubits}
Interestingly, as we will show in the following, there is an ancilla-free joint decomposition of parallel gates. With parallel gates, we refer to gates that can be executed at the same time slice of the circuit, for example, as shown in \cref{fig:parallel_gates} for the case of CNOT gates. 
In \cref{parallel gates}, we start from the protocol in \cref{fig:teleportation_many} and subsequently show that it can be reduced to an ancilla-free decomposition in terms of local operations on the partition $A$ and $B$. Remarkably, again, no classical communication between the partitions is required, and optimality is preserved. The explicit formula for the decomposition is shown in \cref{Final formula}. It contains single-qubit $Z$ gates, as well as multi-qubit operations on all subsets of the $n$ qubits on each partition $A$ and $B$. The operations consist of two different types. The first are multi-qubit rotation gates by angles $\pm\pi/2$. They can be implemented by single-qubit rotation gates sandwiched by ladders of CNOT gates \cite{Cowtan2020}. The second are non-unitary operations, where the single-qubit rotation gates within the CNOT ladders are replaced by computational-basis measurements; see a detailed discussion in \cref{parallel gates}. As a consequence, each term in the decomposition can be implemented with $\mathcal{O}(n)$ CNOT gates. 
Note that an $n$ qubit rotation gate about $\pm \pi/2$ can be implemented with exactly $n$ CNOT gates. Because of that and since the majority of terms in the decomposition of an $n$-qubit rotation gate shown in \cref{Final formula} contain operations on less than $n$ qubits, we expect strong noise reduction by using our scheme. Noise mitigation due to the reduction of costly two-qubit gates in circuit-cutting schemes was recently observed experimentally in Ref.~\cite{Ufrecht2023}.

It has been emphasized that the sampling overhead is not the only relevant metric measuring the quality of a circuit-cutting scheme \cite{Harada2023} but also the number $L$ of different channels to be applied. In principle, the number of experimental shots required to evaluate an expectation value to additive error $\epsilon$ scales as $\mathcal{O}(\gamma^2/\epsilon^2)$ which is independent of $L$. In practice, however, the latency introduced by the additional compilation time needed for each of the $L$ sub-circuits corresponding to the elements $\mathcal{F}_i$ for $i=1,...,L$ on topologically constrained hardware can be the dominant factor in the runtime of the algorithm \cite{Harada2023}. 
As discussed in \cref{Optimal QPD for a pure-sate denisty martrix}, our decomposition guarantees $L=\mathcal{O}(4^n)$ for the case of parallel as well as non-parallel gates, which is an exponential improvement over previous results. It should be noted that we could trade compilation time versus additional circuit depth. Indeed, a multi-qubit rotation gate can be trivially routed to topologically constrained hardware by SWAP insertion. This decreases compilation latency to almost zero but, in turn, might drastically increase the CNOT-gate count and the additional circuit depth $d$, resulting in a more severe impact of noise. Note that a multi-qubit rotation gate can be implemented on fully connected hardware with depth $d=\mathcal{O}(\lceil \mathrm{log}(n) \rceil)$ when replacing each CNOT gate ladder by a balanced tree \cite{Cowtan2020}. For hardware with constrained connections, efficient compilation methods exist for multi-qubit rotation gates based on minimum spanning trees \cite{Gogioso2023} and CNOT gate re-synthesis \cite{Nash2020}.
In what sense the low depth of the circuits can be retained will be the focus of further investigations.

\section{Conclusion}
\label{Outlook and conclusion}
In this work, we introduced joint cutting of non-Clifford two-qubit rotation gates based on virtual teleportation and proved optimality. In the case of parallel gates, we further derived an ancilla-free optimal decomposition. We conclude by stressing the following aspects:
As mentioned before, controlled rotation gates are equivalent to two-qubit rotation gates up to local unitary operations. In addition, CNOT gates are special instances of this gate class. Consequently, the joint virtual teleportation protocol proposed in this work can also be used for optimal joint cutting of CNOT gates with several improvements over previous work \cite{Piveteau2022_circuitcut}. (1) In our virtual teleportation protocol, no classical communication needs to be shared. As a result, no real-time feedback during the quantum computation in the form of local correction operations conditioned on the measurement outcomes on the ancilla qubits is required. (2) As a consequence, the cut circuits can be run in sequence on the same hardware as opposed to parallel execution, which was necessary for previous methods.  (3) By our alternative quasi-probability decomposition of pure-state density matrices, we reach an exponential reduction of the number of terms required compared to Refs.~\cite{Piveteau2022_circuitcut, Vidal_1999}, directly translating into an exponential reduction of the number of channels in the cutting scheme. (4) Our decomposition method allows us to derive ancilla-free optimal joint cutting schemes for CNOT gates, two-qubit rotation gates, and controlled rotation gates, which has not yet been considered in the literature.
\vspace{0.4cm}\\
\noindent
After completion of our manuscript, Refs.~\cite{Schmitt2024,Harrow2024} appeared as preprints which generalize some of our ideas.

\begin{acknowledgements}
C.U.~would like to thank M.~Bechtold, D.~Sutter, and F.~Wagner for helpful discussions.
The research was funded by the project QuaST, supported by the Federal Ministry for Economic Affairs and Climate Action on the basis of a decision by the German Bundestag.
\end{acknowledgements}

\appendix

\section{Optimal QPD for a pure-state density matrix}
\label{Optimal QPD for a pure-sate denisty martrix}
In this appendix, we derive an alternative optimal QPD for arbitrary pure-state density matrices. 
In Ref.~\cite{Vidal_1999}, Vidal and Tarrach consider the following problem: Given two partitions and a bipartite state $\rho$. What is the minimal mixing of the state with a separable state such that the mixed state becomes separable as well? This problem can be rephrased as follows \cite{Piveteau2022_circuitcut}: What is the minimal $R$, called robustness of entanglement, for which 
\begin{equation}
    \rho=(1+R)\eta_1-R\eta_2
\end{equation}
where $\eta_1$ and $\eta_2$ are separable. While such decomposition is hard to find in general, for a pure state $\rho=|\psi\rangle\langle\psi|$ 
the robustness can be calculated via the $m$ Schmidt coefficients $c_j$ with $j=0,...,m-1$ of the $|\psi\rangle$ state as
\begin{equation}
\label{definitionR}
    R=\left(\sum_{j=0}^{m-1}c_j\right)^2-1\,.
\end{equation}
Consequently, 
\begin{equation}
\label{definitiongamma}
    \gamma=1+2R\,.
\end{equation}
In Ref.~\cite{Vidal_1999}, the authors also provide an explicit representation for $\eta_1$ and $\eta_2$ in which, however, the separable states themselves are functions of the Schmidt coefficients.
In contrast, the alternative QPD derived in the following only contains the Schmidt coefficients in the quasi-probabilities. As discussed in the main text, this guarantees that, when applied to the protocol in \cref{fig:virt_teleportation} and \cref{fig:teleportation_many},  we can correct the unwanted measurement results in the classical post-processing step. A similar representation has been derived in Refs.~\cite{Sperling2009a,Sperling2009b}.
In the following, we first derive the general form of the decomposition and subsequently show optimality: 

\noindent
Consider a bipartite state $|\psi\rangle$ on the partitions $A$ and $B$ and an expansion of the form
\begin{equation}
\label{expansion}
    |\psi\rangle =\sum_{j=0}^{m-1} c_j |\varphi_j\rangle \otimes |\varphi_j^\prime\rangle
\end{equation}
with $c_j$ real but not necessarily positive. Then
\begin{align}
    \nonumber
    |\psi\rangle\langle\psi|&=\sum_{j=0}^{m-1} c_j^2 |\varphi_j\rangle \langle\varphi_j| \otimes |\varphi_j^\prime\rangle \langle\varphi_j^\prime|\\
    \label{optimaldecompositionalternative}
    &+2\sum_{i>j}c_i c_j\left( \sigma_{ij}^+-\sigma_{ij}^-\right)
\end{align}
holds.
In this expression
\begin{equation}
\label{definition_sigma}
    \sigma_{ij}^\pm=\frac{1}{\alpha}\sum_{r=1}^\alpha |\xi^\pm_{rij}\rangle \langle \xi^\pm_{rij}|\otimes |\tau_{rij}\rangle \langle \tau_{rij}|\,.
\end{equation}
are separable density matrices with respect to the partitions $A$ and $B$. The states take the form  
\begin{align}
\label{definition xi}
    |\xi^\pm_{rij}\rangle&=\frac{1}{\sqrt{2}}\left(|\varphi_i\rangle\pm \mathrm{e}^{i\phi_r}|\varphi_j\rangle \right)\\
    \intertext{and}
    \label{definition tau}
    |\tau_{rij}\rangle&=\frac{1}{\sqrt{2}}\left(|\varphi_i^\prime\rangle+ \mathrm{e}^{-i\phi_r}|\varphi^\prime_j\rangle \right)\,.
\end{align}
Finally, the phases $\phi_r$ can be chosen such that
\begin{equation}
\label{definition_phase}
    \sum_{r=1}^\alpha \mathrm{e}^{i\phi_r}=    \sum_{r=1}^\alpha \mathrm{e}^{i2\phi_r}=0\,.
\end{equation}
Furthermore, if \cref{expansion} is the Schmidt decomposition of $|\psi\rangle$ - in this case $c_j\geq0$ -  then \cref{optimaldecompositionalternative}  is optimal in that it achieves the minimal possible $\gamma$.\\

\noindent
\textit{Proof}:
The statement is most easily proven by substituting Eqs.~(\ref{definition_sigma})-(\ref{definition_phase})  into \cref{optimaldecompositionalternative}. It remains to show optimality in case \cref{expansion} is the Schmidt decomposition of $|\psi\rangle$. To this end, we sum over the absolute of the coefficients for all terms with the result stated in \cref{definitionR} and \cref{definitiongamma}. Therefore, the decomposition is optimal.

We choose $\phi_r=2\pi r/\alpha$ for integer $\alpha\geq3$. The choice $\alpha=3$ involves the least amount of terms, but we observe in \cref{parallel gates} that $\alpha=4$ allows removing ancilla qubits for parallel gates. \cref{optimaldecompositionalternative} involves $\mathcal{O}(m^2)$ terms. This is an exponential improvement to the decomposition used in Ref.~\cite{Vidal_1999} with $\mathcal{O}(2^m)$ terms.

\section{Lower bounds}
\label{Lower bounds}
 In this appendix we prove a lower bound on  $\gamma_\mathrm{joint}^{(n)}$, the $\gamma$ parameter for jointly cutting $n$ two-qubit rotation gates. This bound will agree with the upper bound established in \cref{derivation_gamma_3} by the joint virtual teleportation scheme, thereby proving optimality.
 A lower bound on the $\gamma$ parameter of a quantum gate can be obtained from the QPD of the Choi state of the gate's unitary \cite{Piveteau2022_circuitcut}. The argument is the following: Since the Choi state of a gate's unitary can be formed by local (with respect to the partitions $A$ and $B$) operations, the $\gamma$ parameter of the gate cannot possibly be smaller than that of the Choi state. If it were, construction of the Choi state with smaller $\gamma$ would be possible, leading to a contradiction. Thus, \cref{calculation gamma} provides
\begin{equation}
\label{lower bound}
    \gamma\geq 2\Big(\sum_j c_j\Big)^2-1
\end{equation} 
where $c_j$ are the Schmidt coefficients of the Choi state of the gate. Further
note that the Choi state of a gate whose matrix representation is diagonal in the computational basis has the same Schmidt coefficients as the normalized vector containing the diagonal elements. In our case, this vector is just $|\Psi_{k_1l_1...k_nl_n}\rangle$ for $k_s=l_s=0$ for all $s$ up to local operations. Consequently, the lower bound matches the upper, proving the optimality of the joint virtual teleportation approach for joint cutting of two-qubit rotation gates. 

Ref.~\cite{Piveteau2022_circuitcut} proved optimality for a large class of gate decomposition derived in Ref.~\cite{Mitarai2021b}.
Note, however, that the lower bound obtained in this way can be relatively loose for some gates. E.g.,~in case of a $K$-qubit version of a Toffoli gate, it can be shown that the right-hand side of \cref{lower bound} for a cut after one qubit tends to one as $K\rightarrow \infty$.
Since, however, this gate can be used to create a Bell state between two partitions, $\gamma\geq3$ must be true.

\section{Parallel gates}
\label{parallel gates}
In this appendix, we derive the ancilla-free decomposition of $n$ parallel two-qubit rotation gates for which we start from the protocol in \cref{fig:teleportation_many}. First, we define the map
\begin{equation}
\label{Map}
    \Lambda(|\psi\rangle \langle \psi|)[\cdot]=2^n\sum_{i,j}\langle i|\psi\rangle \langle \psi |j\rangle P_i \,\cdot\,P_j
\end{equation}
where $P_i=|i\rangle\langle i|$ and $\{|i\rangle\}_{i=0}^{2^n-1}$ is the computational basis. 
This map corresponds to the transformation $U \cdot U^\dagger$ where $U$ contains the computational basis elements of $|\psi\rangle$ multiplied by $\sqrt{2^n}$ on its diagonal. Thus, $U$ is unitary if  $|\langle i |\psi\rangle|=1/\sqrt{2^n}$ for all $i$. Further, note that $\Lambda$ is linear in its first argument, and if $|\psi\rangle$ factorizes over two partitions, then the map does as well.

Let us define as $\mathcal{R}_{zz}^{(n)}$ the $2n$-qubit unitary channel corresponding to the $n$ two-qubit rotation gates we want to cut and a state $|\psi\rangle$ such that
\begin{equation}
\label{channelZ}
  \Lambda(|\psi\rangle \langle \psi|)[\cdot] = \mathcal{R}_{zz}^{(n)}\,.
\end{equation}
With the Hadamard gate $H$ and the phase gate $S$ this state is
\begin{equation}
    |\psi\rangle=H^{\otimes n}\otimes (H S^\dagger)^{\otimes n}|\Psi\rangle
\end{equation}
with $|\Psi\rangle$ as in \cref{decomposition_n_state} but with $k_s=l_s=0$ for all $s$, that is
\begin{equation}
\label{decomposition_n_state_kl0}
|\Psi\rangle=\sum_{j\in\{0,1\}^n}c_j |j_1...j_n\rangle| j_n...j_1\rangle
\end{equation}
with $c_j=\Pi_{s=1}^n\mathscr{c}_{j_s}$
and 
\begin{equation}
\mathscr{c}_{j_s} = \left\{\begin{array}{lr}
        \mathrm{cos}(\theta_s/2) & \quad\text{for } j_s=0\\
        \mathrm{sin}(\theta_s/2) &\quad \text{for } j_s=1
        \end{array}\right.\,. 
\end{equation}

It is left to insert a QPD of $|\psi\rangle\langle\psi|$ calculated with \cref{Optimal QPD for a pure-sate denisty martrix} into  \cref{Map} and then use linearity and factorization properties of $\Lambda$. Comparison with \cref{expansion} shows $m=2^n$, $|\varphi_j\rangle=H^{\otimes n}|j_1,...,j_n\rangle$, and $|\varphi_j^\prime \rangle=(HS^\dagger)^{\otimes n}|j_n,...,j_1\rangle$. To calculate the map $\Lambda$, we first make use of linearity and calculate the different terms individually.
When we define the unitary channel
\begin{equation}
\mathcal{Z}_j[\cdot]=Z^{j_1}\otimes ...\otimes Z^{j_n}\cdot Z^{j_1}\otimes ...\otimes Z^{j_n}
\end{equation}
 corresponding to the application of single qubit $Z$ gates, we find
\begin{equation}
    \Lambda(|\varphi_j\rangle\langle\varphi_j|)[\cdot]=\mathcal{Z}_j[\cdot]
\end{equation}
where we used
\begin{equation}
    \sqrt{2}\sum_{k_s\in\{0,1\}}\langle k_s|H|j_s\rangle P_{k_s}=\sum_{k_s\in\{0,1\}}(-1)^{j_sk_s} P_{k_s}=Z^{j_s}\,.
\end{equation} 
Equally, we find the same result for $|\varphi_j^\prime\rangle$ for $j $ replaced by $\tilde{j}:=(j_n,...,j_1)$. The evaluation of the states $|\xi^\pm_{rij}\rangle$ and $|\tau_{rij}\rangle$ requires a little bit more work. First, we insert 
 $|\xi^\pm_{rij}\rangle$, defined in \cref{definition xi}, with the result
\begin{align}
\nonumber
    \sqrt{2}^n\sum_k\langle k&|\xi^\pm_{rij}\rangle P_k=Z^{i_1}\otimes...\otimes Z^{i_n}\times\\
    \label{write as multi-qubit_2}
    &\left[(\mathbb{I}^{\otimes n}\pm\mathrm{e}^{i\phi_r} Z^{j_1-i_1}\otimes...\otimes Z^{j_n-i_n})/\sqrt{2}\right]\,.
\end{align}
For  $\phi_r=2\pi r/\alpha$ with $\alpha=4$, we first consider  $\mathrm{exp}(i\phi_1)=i$ and $\mathrm{exp}(i\phi_3)=-i$.
In this case, the expression in the brackets is unitary. Indeed, 
\begin{equation}
    R_{ij}(\pm\pi/2)=\frac{\mathbb{I}^{\otimes n}\mp iZ^{j_1-i_1}\otimes...\otimes Z^{j_n-i_n} }{\sqrt{2}}
\end{equation}
is a multi-qubit rotation gate on the qubits with $|i_s-j_s|=1$. Again, we define the corresponding channel $\mathcal{R}_{ij}(\pm\pi/2)$ using calligraphic notation. Furthermore, we introduce the abbreviation 
\begin{equation}
    \mathcal{R}_{ij}=\frac{\mathcal{R}_{ij}(\pi/2)-\mathcal{R}_{ij}(-\pi/2)}{2}
\end{equation}
of the unitary channels.
We now turn to \cref{write as multi-qubit_2} with $\mathrm{exp}(i\phi_2)=-1$ and $\mathrm{exp}(i\phi_4)=1$. In this case, we define
\begin{equation}
P^k_{ij}=\frac{\mathbb{I}^{\otimes n}+(-1)^k Z^{j_1-i_1}\otimes...\otimes Z^{j_n-i_n} }{2}
\end{equation}
for $k=0,1$, which is non-unitary. The channel $\mathcal{P}_{ij}^k$ can be evaluated on a quantum computer using a ladder of CNOT gates \cite{Mitarai2019} and a projector on $|0\rangle$ or $|1\rangle$, e.g.
\begin{equation}
\label{CNOT_chain}
\frac{\mathbb{I}^{\otimes t}+(-1)^k Z^{\otimes t}}{2}\quad=\quad\vcenter{\hbox{\includegraphics[width=107pt]{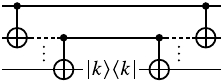}}}
\end{equation}
for integer $t>0$.
 We then define 
 \begin{equation}
 \label{measurement operation}
     \mathcal{P}_{ij}=\mathcal{P}_{ij}^0-\mathcal{P}_{ij}^1
 \end{equation}
which is neither trace-preserving nor positive. However, $\mathcal{P}_{ij}$ is the difference of two completely positive trace non-increasing (CPTN) maps, whose sum is completely positive and trace preserving (CPTP). Such maps can be simulated on a quantum device without extra sampling overhead \cite{Ufrecht2023, Piveteau2022_circuitcut, Mitarai2021,Mitarai2021b}. Thus, both $\mathcal{R}_{ij}$ and $\mathcal{P}_{ij}$ can be simulated with $\gamma=1$.
 The evaluation of  \cref{Map} for $|\tau_{rij}\rangle$ proceeds along the same lines. We only need to replace $i,j\rightarrow \tilde{i}, \tilde{j}$, choose the plus sign in \cref{write as multi-qubit_2}, and replace $\phi_r\rightarrow -\phi_r$.
Due to the extra $S^\dagger$ gate in the definition 
of $|\varphi^\prime\rangle$, we additionally
have to replace $Z\rightarrow -iZ$, introducing a factor $(-i)^{\nu_{ij}}$
 with
 \begin{equation}
     \nu_{ij}=\sum_{s=1}^n (j_s-i_s)\,.
 \end{equation}
Collecting all the terms, we finally arrive at 
 \begin{widetext}
\begin{equation}
\label{Final formula}
    \mathcal{R}_{zz}^{(n)}=\sum_{j \in\{0,1\}^n}\! c_j^2\, \mathcal{Z}_j\otimes \mathcal{Z}_{\tilde{j}}
     +2\sum_{i>j}c_i c_j[\mathcal{Z}_i\otimes \mathcal{Z}_{\tilde{i}}]\circ
     \begin{cases}
\;\,(-1)^{\nu_{ij}/2}\,\Big[
    \mathcal{P}_{ij}\otimes\mathcal{P}_{\tilde{i}\tilde{j}}-\mathcal{R}_{ij}\otimes \mathcal{R}_{\tilde{i}\tilde{j}}\Big]\quad &\text{for $\nu_{ij}$ even}\\
    \;\,(-1)^{(\nu_{ij}-1)/2}\Big[\mathcal{R}_{ij}\otimes \mathcal{P}_{\tilde{i}\tilde{j}}
    +\mathcal{P}_{ij}\otimes\mathcal{R}_{\tilde{i}\tilde{j}}\Big] \quad &\text{for $\nu_{ij}$ odd}
     \end{cases}\,.
\end{equation}
 \end{widetext}
This decomposition is optimal with
\begin{equation}
    \gamma=2\,\Pi_{s=1}^n(1+|\mathrm{sin}(\theta_s)|)-1\,.
\end{equation}
It consists of single-qubit $Z$
gates, multi-qubit $Z$ rotations by the angles $\pm\pi/2$ and measurements within CNOT-gate ladders. 
Due to the double sum, the number of terms in \cref{Final formula} grows as $\mathcal{O}(4^n)$. Interestingly, the decomposition achieves the optimal $\gamma$ parameter even though no exchange of classical information is required between the partitions.

\end{document}